# 2-gram-based Phonetic Feature Generation
# for Convolutional Neural Network in Assessment of Trademark Similarity

Kyung Pyo Ko*[1], Kwang Hee Lee*[2], Mi So Jang[1], Gun Hong Park[1]

Sejin Mind[1], Artificial Intelligence Research Institute[2]
kkp1008@snu.ac.kr, kwanghee@airi.kr,
jangms@mju.ac.kr, ghpark@sejinmind.com
* indicates equal contribution

*Abstract*

*A trademark is a mark used to identify various commodities. If same or similar trademark is registered for the same or similar commodity, the purchaser of the goods may be confused. Therefore, in the process of trademark registration examination, the examiner judges whether the trademark is the same or similar to the other applied or registered trademarks. The confusion in trademarks is based on the visual, phonetic or conceptual similarity of the marks. In this paper, we focus specifically on the phonetic similarity between trademarks. We propose a method to generate 2D phonetic feature for convolutional neural network in assessment of trademark similarity. This proposed algorithm is tested with 12,553 trademark phonetic similar pairs and 34,020 trademark phonetic non-similar pairs from 2010 to 2016. As a result, we have obtained approximately 92% judgment accuracy.*

***keywords*** *: Phonetic similarity, Trademark, N-gram, Romanization, International phonetic alphabet,*
              *Convolutional neural network*

## 1. Introduction

According to the trademark manual provided by the Office for Harmonization in the Internal Market(OHIM), a trademark is a mark for identifying one's goods or service. A trademark has the form of text, image, sound, and smell, and depending on its value, becomes an intangible asset for an enterprise. Therefore trademarks are very important as intellectual property and should be legally protected. Trademark registration provides a legal right to enterprises, products, and services of a subject using the trademark and is a means to prevent others from misusing.

The number of trademark applications has risen 15.3% globally over the previous year to about six million by 2015. The trend is steadily increasing. According to the United State Patent and Trademark Office(USPTO), the number of trademark applications in 2017 was 440,768, 13.4% more than the previous year's 389,000 trademark applications. It is expected that the number of trademark applications will continue to increase in the future due to the addition of various types of trademarks and the increase of industrial groups. In the registration examination of a trademark, the examiner confirms whether the trademark is likely to infringe on another existing trademark, and confirms the similarity of the trademark's appearance, pronunciation of title and conceptual thinking in this process. In the case of trademark phonetic similarity judgments in trademarks, the trademark examiner's judgment was considered on the basis of confusion of the public using the trademark. However, this judgement was subjective and inconsistent. In this paper, we propose a new method to judge the phonetic similarity between trademark pair.

In Section 2 of this paper, we analyze the phonetic similarity of titles and explain the proposed algorithm. In Sections 3, 4 and 5, we discuss the proposed algorithm in more detail, and we analyze and discuss the experimental results.

## 2. Related Work

The phonetic similarity measurement of trademarks means to compare two words or two sentences numerically. Thus, there have been studies that develop words into phonetic features.

For example, Soundex[11] was developed by Odell and Russell and patented in 1918. Soundex is an algorithm for indexing the pronunciation of a word, and uses a consonant and a number of words to map a word to a specific index value. For example, if the words 'Smith' and 'Smythe' are converted by Soundex, the same value is returned to S530.

Metaphone[12] is an improved algorithm for Soundex and was invented by Lawrence Philips in 1990. Like Soundex, Metaphone shares the same index key with similar pronunciation words, and uses information about variations and discrepancies to generate a more accurate encoding than Soundex. Metaphone has been improved to encode for various pronunciations such as Double Metaphone and Metaphone 3.

One of the phonetic similarity algorithms that solves the technical limitations of Soundex is the alignment of phonetic sequences(ALINE) algorithm developed by Kondrak in 2000[6]. The ALINE algorithm, developed for use in a comparative paper of the similarity of drug names, expresses the phonemes of word sequences as feature vectors. Each vector consists of ten binary features and two multivalued features. With these features, the ALINE algorithm specifies the similarity score of each phoneme pair based on the weight analysis of consonants and vowels. The advantage of the ALINE algorithm is that Soundex indexes words into four limited numbers, but the ALINE algorithm does not have to cut the words into a limited number. Metaphone also removes vowels and indexes vowels with consonants only. However, the ALINE algorithm uses vowels to match two words without removing the vowels. The ALINE algorithm uses the phoneme instead of indexing the phoneme with a specific value in Soundex or Metaphone. In this way, the similarity is calculated by aligning the two words with similar pronunciation parts.

## 3. N-gram based Phonetic Feature Generation

In this paper, we propose an n-gram-based Phonetic Feature (PF) generation algorithm for trademark similarity assessment. The two PFs generated from trademark pair are used as an input of the convolutional neural network. The convolutional neural network outputs the phonetic similarity score between trademark pair. Our proposed method is summarized as follows (see Figure 1.).

(1) The multilingual trademark is converted into its phonetic transcription using International Phonetic Alphabet (IPA) (Section 3.1).
(2) The IPA word is segmented into consecutive n-grams (Section 3.2).
(3) In order to generate PF, these n-grams are connected by straight lines in n-D space by pronunciation order (Section 3.3).
(4) The PF pairs are trained by our convolutional neural network for assessment of trademark similarity (Section 3.4).

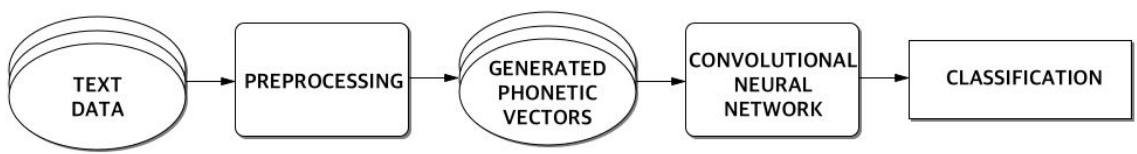

**Figure 1**. Workflow of Proposed Method

## 3.1 Preprocessing

In this step, the trademark words are converted into Romanization or IPA using rule-based converting softwares [8, 9]. Table 1 and 2 show examples of both cases (Table 1, 2).

|  | Sample 1 | Sample 2 | Sample 3 |
|---|---|---|---|
| Korean | 아디다스 | 구글 | 페이스북 |
| Romanization | adidaseu | gugeul | peiseubug |

**Table 1**. Korean to Romanization

|  | Sample 1 | Sample 2 | Sample 3 |
|---|---|---|---|
| English | Adidas | Google | Facebook |
| IPA | ədidəs | gugəl | fesbʊk |

**Table 2**. English to IPA

## 3.2 N-gram-based Word Segmentation

We extract n-gram segments from a word transformed into IPA(or Romanization) symbols. An IPA word of length $N$ can be represented by a sequence of IPA symbols $S = (s_1, s_2, \ldots, s_{N+n-1})$. An n-gram of the sequence $S$ is any n-long subsequence of consecutive IPA symbols. The $i^{th}$ n-gram of S is defined as $G_i = (s_i, s_{i+1}, \ldots, s_{i+n-1})$. In this paper, we use 2-gram byte-level to generate two-dimensional (2D) PF (u: width, v: height) from the IPA word. For example, the string "TEXT" would be composed of the following 2-gram byte-level: -T, TE, EX, XT, T_. The hyphen character ("-") and underscore character ("_") are used here to represent the beginning and ending of the string. In this case, a sequence of IPA symbol $S$ of the word "TEXT" is represented as

$$S = (s_1, s_2, s_3, s_4, s_5, s_6) = ("-", "T", "E", "X", "T", "\_") \quad (1)$$

and the first n-gram $G_1$ is:

$$G_1 = (s_1, s_2) = ("-", "T") \quad (2)$$

## 3.3 Phonetic Feature Generation using 2-Gram Segments

We propose 2D PF generation method based on aforementioned 2-gram segments. The generated PF is used as input of our convolutional neural network. In order to obtain the PF $\in R^{u \times v}$ of width $u$ and height $v$, we first determine a representative value for each 2-gram $G_i$, which is corresponding to a 2D coordinate of the PF. Both coordinate axes of the PF are composed of identical 43 IPAs, starting("-") and ending("_") symbols (totally 45 symbols). We set (u, v) to (128, 128). To assign each symbol into a value between 0 to 127, we make a dictionary. The dictionary has 45 pairs, which are composed of the symbol and corresponding value.

According to official IPA Table(revised to 2005)[10], the vowels and consonants are represented separately. And we figure out that symbols with similar pronunciation are bound together. So we group phonetically similar symbols. In order to reflect the difference in sound between the groups, we narrow the interval in the same group, and widen the interval with other groups. The mapping value for each symbol was determined experimentally and the dictionary is shown in Figure 2.

```
{"-": 16, "i": 19, "ɪ": 21, "y": 23, "e": 25, "ɛ": 27,
"ə": 29, "ɜ": 30, "æ": 32, "a": 33, "ɑ": 35, "ʌ": 37,
"ɔ": 38, "o": 40, "ʊ": 42, "w": 44, "u": 46, "h": 48,
"p": 51, "b": 53, "v": 55, "f": 57, "c": 59, "k": 61,
"q": 63, "g": 66, "d": 69, "t": 72, "θ": 74, "ð": 76,
"s": 79, "ʃ": 81, "ʧ": 83, "x": 85, "z": 88, "ʒ": 90,
"ʤ": 92, "j": 94, "r": 97, "ɹ": 99, "l": 101, "m":
104, "n": 106, "ŋ": 108, "_": 111}
```

**Figure 2**. IPA with an Dictionary

Table 3 and 4 show the mapping examples transformed by the dictionary.

| Trademark | Romanize | Mapping List |
|---|---|---|
| 아디다스 | -adidaseu_ | [16, 33, 69, 19, 69, 33, 79, 25, 46, 111] |
| 구글 | -gugeul_ | [16, 66, 46, 66, 25, 46, 101, 111] |
| 페이스북 | -peiseubug_ | [16, 51, 25, 19, 79, 25, 46, 53, 46, 66, 111] |

**Table 3**. Mapping Examples Transformed by the Dictionary from Romanization

| Trademark | IPA | Mapping List |
|---|---|---|
| Adidas | -ədidəs_ | [16, 29, 69, 19, 69, 29, 79, 111] |
| Google | -gugəl_ | [16, 66, 46, 66, 29, 101, 111] |
| Facebook | -fesbʊk_ | [16, 57, 25, 79, 53, 42, 61, 111] |

**Table 4**. Mapping Examples Transformed by the Dictionary from IPA

The representative value for $G_i$ defined as

$$PF(G_i) = Z \prod_{k=0}^{i} \gamma^k \quad (3)$$

where $Z$ is a scale factor and $\gamma$ is a discount factor.
In the next step, the coordinates of all the $G_i$s are connected in straight lines in the pronunciation order ($G_1 \rightarrow G_2 \rightarrow \ldots \rightarrow G_N$). Intensity of the each $i^{th}$ straight line is assigned the representative value of $G_i$ and thickness of the straight line is normalized to the sum of the lengths of the straight lines.

For example, if the trademark of Section 3.1 is divided based on the 2-gram partitioning coordinates and corresponds to the x-axis and the y-axis, the following coordinate values Xn and Yn, can be obtained.

| Example | Xn, Yn |
|---|---|
| -ədidəs_ | [(16, 29), (29, 69), (69, 19), (19, 69), (69, 29), (29, 79), (79, 111)] |

Table 5. Example of X Y Coordinates by Using 2-gram

By plotting the points corresponding to the above coordinate values and connecting the lines, the following feature vector values can be obtained (Figure 3).

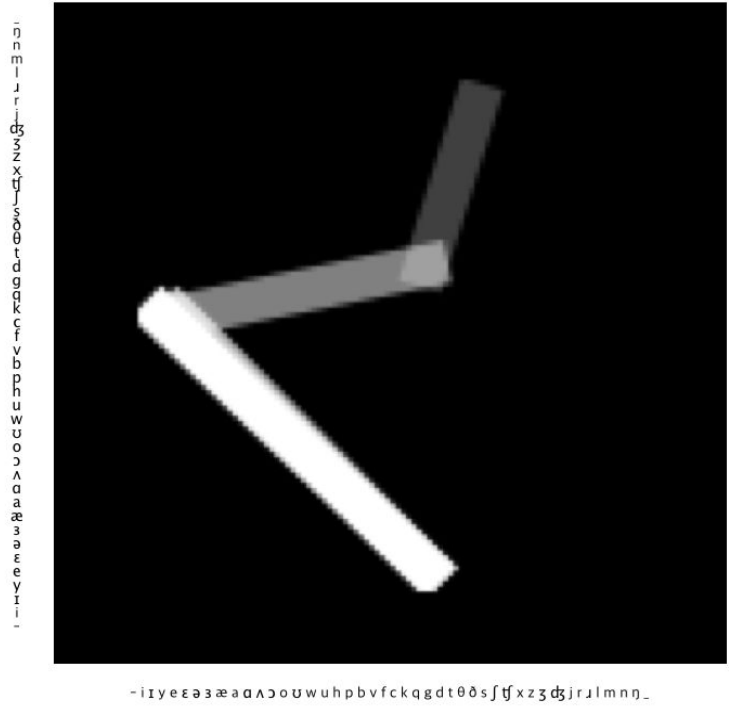

Figure 3. Phonetic Features of ADIDAS

| Trademark 1 | Phonetic Features | Trademark 2 | Phonetic Features |
|---|---|---|---|
| 아디다스 | | Adidas | |
| 구글 | | Google | |
| 페이스북 | | Facebook | |

Table 6. Phonetic Features

## 3.4 Convolutional Neural Network for Assessment of Trademark Similarity

In order to classify phonetic similarity of trademark pair, we use the convolutional neural network model as shown Figure 4. The PFs generated from two trademarks are combined into two channel image as input of CNN. To visualize, third channel image is filled by zeros (Table 6,7 and 8).

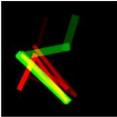

Table 7. Phonetic Features of Trademark Pairs

As shown in Figure 3, the overlapping part of the red channel and the green channel is shown in yellow. We can see that there are many overlapping parts because similar images are displayed in the case of a trademark having similar pronunciation. The following trademarks are typical cases in which the registration of a trademark is rejected due to similar pronunciation (Table 7). The proposed method is applied.

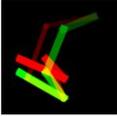

Table 8. Similar Representative Phonetic Samples

The ALINE algorithm is characterized by arranging similar parts in two words. These are the results of the ALINE algorithm and proposed method.

| Input | ALINE Algorithm | Proposed Algorithm |
|---|---|---|
| grass : gramen | \|\| g r æ \|\| s <br> \|\| g r a \|\| men | 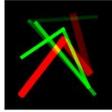 |
| three : tres | \|\| θ r iy \|\| <br> \|\| t r e \|\| s | 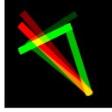 |
| blow : flare | \|\| b l o \|\| w <br> \|\| f l a \|\| re | 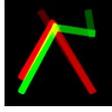 |
| full : plenus | \|\| f u l \|\| <br> \|\| p - l \|\| enus | 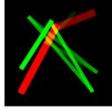 |
| fish : piscis | \|\| f i s \|\| <br> \|\| p i s \|\| kis | 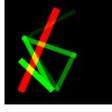 |
| tooth : dentis | \|\| t uw θ \|\| <br> den \|\| t i s \|\| | 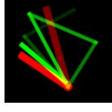 |

**Table 9**. Results of ALINE and Proposed Algorithm

Through the above preprocessing process, the data is trained through the convolutional neural network as shown in Figure 4 using the PF values of the words. The following model uses a batch to pass 128 x 128 input values to a convolutional neural network. After passing through two convolutional neural network layers and passing through two fully connected neural network layers, a Softmax function is used to output the more probable results.

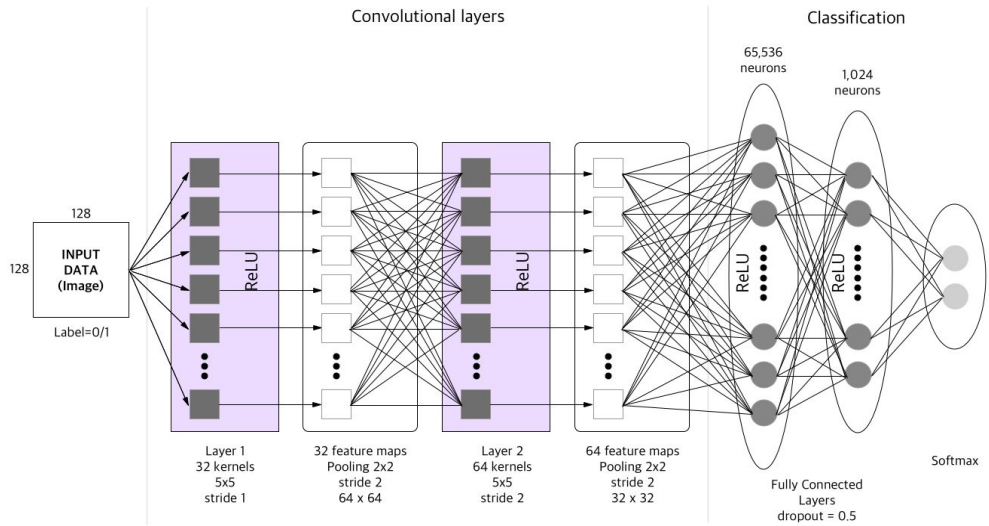

**Figure 4**. Convolutional Neural Network Model

## 4. Experiments

### 4.1 Dataset

For this paper, we have rejected trademark data caused by phonetic similarity at the Korea Intellectual Property Rights Information Service(KIPRIS). From 2010 to 2016, 12,553 pairs of phonetically similar brand data were rejected. Thus, 34,020 pairs of phonetically non-similar brand data are used for the experiment. Figures 5 and 6 are visual representations of the cosine distance of each pair of trademarks. In the case of similar pairs, the distribution is shifted closer to 0 than the data distribution of non-similar pairs.

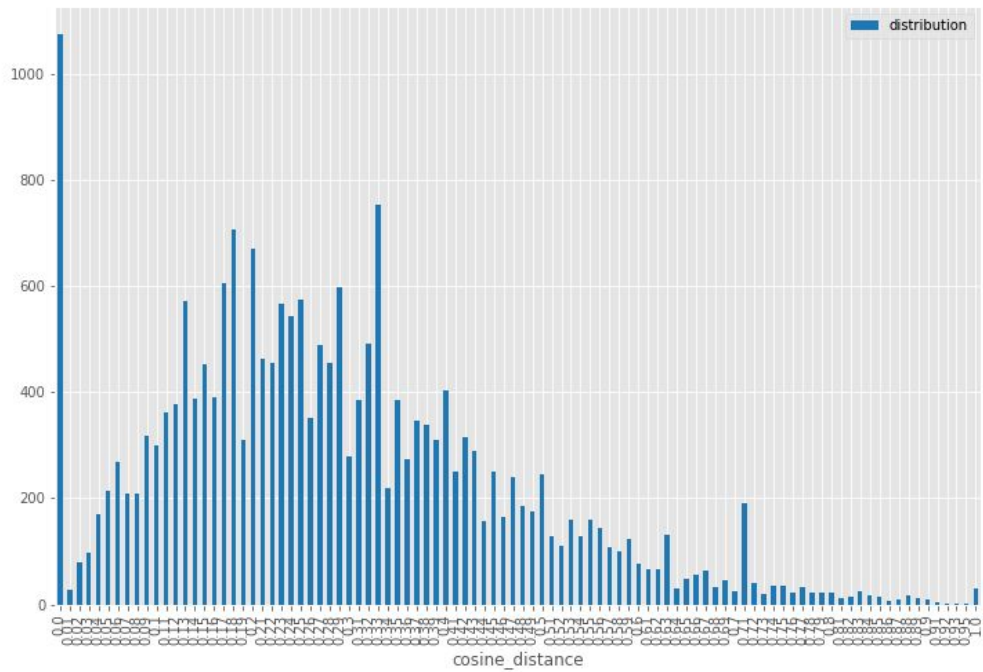

**Figure 5**. Cosine Distance Distributions at Similar Trademark Pairs

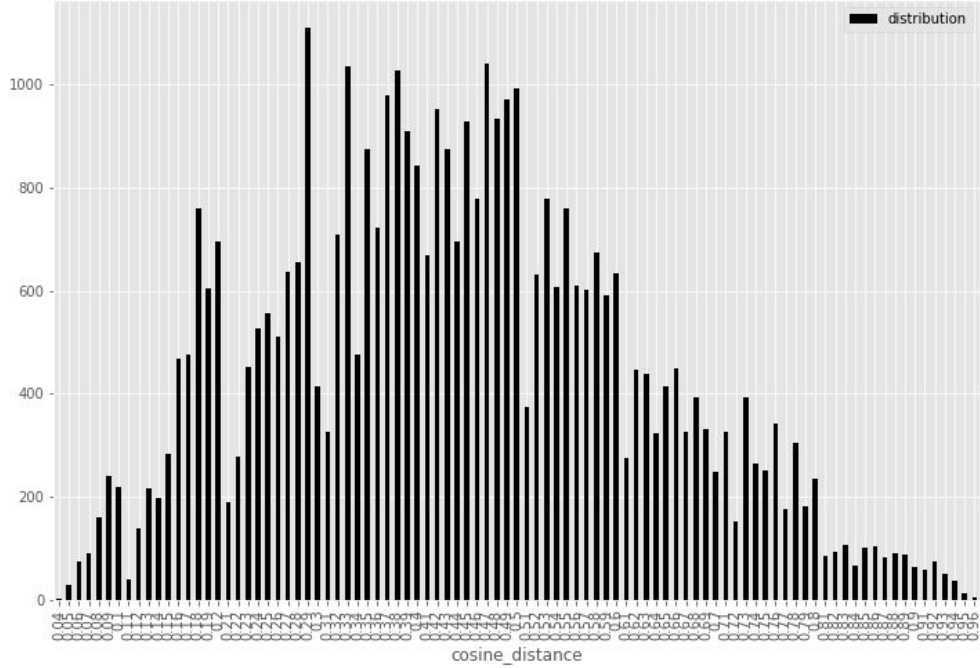

**Figure 6**. Cosine Distance Distributions at Non-Similar Trademark Pairs

### 4.2 Implementation Details

(1) Parameters of Equation

$$PF(G_i) = Z \prod_{k=0}^{i} \gamma^k$$

In the above equation, Z = 255 and γ = 0.9 are set as initial values.

(2) Hyperparameters of the deep neural network

Conv2d_1 :
kernel=(5, 5, 32), padding="same", stride=1
MaxPool_1 :
kernel=(2, 2), stride=2
Conv2d_2 :
kernel=(5, 5, 64), padding="same", stride=1
MaxPool_2 :
kernel=(2, 2), stride=2
Activation Function = Relu
Optimizer = Adam
Dropout = 0.5
Tensorflow versions = 1.4

### 4.3 Evaluation

As a result of evaluation, we get 0.927 accuracy in validation dataset. The training data and validation data are split in a 9:1 ratio. For comparing with proposed method, we get accuracy up to 0.742 in cosine distance algorithm.

| Algorithm | Cosine Algorithm | Proposed Algorithm |
|---|---|---|
| Accuracy | 0.742 | 0.927 |

**Table 10**. Result of Evaluation

## 5. Conclusions

In this paper, we proposed novel method for n-gram-based phonetic feature generation. And we can get meaningful result from this feature with basic CNN model. As a result of using the proposed method to judge the similarity of the trademark title, an accuracy of 0.927 was obtained in the validation dataset. This is about 24.9% higher than the cosine distance algorithm.

**Acknowledgement**

This work was supported by Institute for Information & communications Technology Promotion(IITP) grant funded by the Korea government(MSIT) (No.2017-0-01181, Deep Learning Based Text Trademark Search Service & No.2017-0-01778, Development of Explainable Human-level Deep Machine Learning Inference Framework)